\documentclass[12pt]{article}

\begin{document}

\begin{center}
{\Large\bf{}The modification of the Bel-Robinson energy-momentum}
\end{center}

\begin{center}
Lau Loi So{\footnote{email address: s0242010@gmail.com}} 
\end{center}

\begin{abstract}
For describing the non-negative gravitational energy-momentum in
terms of a pure Bel-Robinson `momentum' in a quasi-local small
sphere limit, the Bel-Robinson tensor $B$ is desirable. However,
we found this Bel-Robinson `momentum' can be modified such that it
still satisfy the non-spacelike and future pointing requirement.
These particular energy-momentum properties can be obtained from a
linear combination between $B$ with other tensor $S$ in a small
sphere limit. This implies that the Landau-Lifshitz pseudo-tensor
is no longer disqualified for this non-spacelike and future
pointing requirement. Moreover, we constructed a certain linear
combination using tensors $B,S,T$ that gives the dominate energy
condition in a small sphere region.
\end{abstract}

\section{Introduction}
Bel and Robinson proposed an energy-momentum 4-index tensor since
1958~\cite{Bel-1,Bel-2,Robinson-1,Robinson-2}. This tensor is
constructed from the Weyl tensor that analogy with the
electromagnetic stress tensor. The Bel-Robinson tensor $B$ is
desirable for describing the non-negative gravitational energy in
a small sphere limit. According to general relativity, there does
not exist unique definition of the local energy of the
gravitational field. The Bel-Robinson tensor tensor is desirable
for this local energy since the Ricci tensor vanishes in
vacuum~\cite{Horowitz-small-sphere}.

It is known, similar with the electromagnetic stress tensor, the
Bel-Robinson tensor possesses many nice properties: completely
symmetric, divergence free and trace free. Moreover, it satisfies
the dominant energy condition~\cite{Senovilla}. Dominant energy
condition automatically fulfills the future pointing and
non-spacelike properties, but the converse is not guaranteed.
According to the Living Review article, Szabados said (see 4.2.2
in~\cite{Szabados}): ``Therefore, in vacuum in the leading $r^{5}$
order any coordinate and Lorentz-covariant quasi-local
energy-momentum expression which is non-spacelike and future
pointing must be proportional to the Bel-Robinson `momentum'
$B_{\mu\lambda\xi\kappa}t^{\lambda}t^{\xi}t^{\kappa}$.''  Note
that here $t^{\alpha}$ is the timelike unit vector and the
referred `momentum' means 4-momentum.

Previously, it is believed that the Bel-Robinson `momentum' was a
natural choice and unique choice for describing the non-negative
gravitational quasi-local energy-momentum expression. In the past,
we thought there were only two gravitational energy-momentum
expressions contribute the desirable positive definite energy
since they give a positive multiple of the Bel-Robinson `momentum'
in a small sphere limit. They are the Papapertrou
pseudo-tensor~\cite{Papapetrou,SoNesterCQG2009,1stSoCQG2009} and
tetrad-teleparallel energy-momentum gauge current
expression~\cite{Pereira,CJP}. We even confidently concluded that
the Landau-Lifshitz pseudo-tensors cannot warranty the positivity,
but now we claim it was mistaken~\cite{1stSoCQG2009}.

Basically, quasi-local methods are not fundamentally different
than pseudo-tensor methods~\cite{NesterPRL,Boot}.  Although
pseudo-tensor is an co-ordinates dependent object, it is still a
practical way to calculate the work done for an isolated system
from an external universe, e.g., tidal heating through
transferring the gravitational field from Jupiter to its satellite
Io. More concretely, tidal heating is a real physical observable
irreversible process that Jupiter distorts and heats up Io, Purdue
used the Landau-Lifshitz pseudotensor to calculate the tidal
heating for Io in 1999~\cite{Smith,Thorne,Purdue}. Positive
gravitational energy is required for the stability of the
spacetime~\cite{Horowitz} and any quasi-local stress expression
which gives the Bel-Robinson `momentum' is the desirable
candidate. Moreover, evaluating the quasi-local energy-momentum
around a closed 2-surface, we can use the Bel-Robinson `momentum'
to test whether the expression can have a chance to give the
positivity at the large scale or not. Since negative quasi-local
energy guarantees negative for a large scale, while positive
quasi-local energy might have a chance for the large scale.
Checking the result for the gravitational energy in a small region
is an economic way because the positivity energy proof is not
easy.

The motivation for reviewing the argument that raised by Szabados
is that we suspect there may exist a relaxation such that the
desirable physical requirements can be satisfied, i.e., future
pointing and non-spacelike. We claim that the verification and
explanation given by Szabados is necessary but not sufficient. For
example, we find the energy-momentum of the Landau-Lifshitz
pseudo-tensor does satisfy the future pointing and non-spacelike
requirement, though not a multiple of the pure Bel-Robinson
`momentum' in a small sphere limit. Moreover, We claim the
Bel-Robinson tensor lost its privilege to achieve the dominate
energy condition in a small sphere region, because a certain
linear combination of the energy-momentum expression between
$B,S,T$ gives the same condition.

\section{Technical background}
Making use a Taylor series expansion, the metric tensor can be
written as
\begin{equation}
g_{\alpha\beta}(x)
=g_{\alpha\beta}(0)+\partial_{\mu}g_{\alpha\beta}(0)x^{\mu}
+\frac{1}{2}\partial^{2}_{\mu\nu}g_{\alpha\beta}(0)x^{\mu}x^{\nu}+\ldots.
\end{equation}
At the origin in Riemann normal coordinates
\begin{eqnarray}
g_{\alpha\beta}(0)&=&\eta_{\alpha\beta},
\quad\quad\quad\quad\quad\quad\quad\quad~~
\partial_{\mu}g_{\alpha\beta}(0)=0, \\
-3\partial^{2}_{\mu\nu}g_{\alpha\beta}(0)
&=&R_{\alpha\mu\beta\nu}+R_{\alpha\nu\beta\mu},\quad\quad
-3\partial_{\nu}\Gamma^{\mu}{}_{\alpha\beta}(0)
=R^{\mu}{}_{\alpha\beta\nu}+R^{\mu}{}_{\beta\alpha\nu}.
\end{eqnarray}
In vacuum the Bel-Robinson tensor $B$, and tensors $S$ and $T$ are
defined as follows
\begin{eqnarray}
B_{\alpha\beta\mu\nu}&:=&R_{\alpha\xi\mu\kappa}R_{\beta}{}^{\xi}{}_{\nu}{}^{\kappa}
+R_{\alpha\xi\nu\kappa}R_{\beta}{}^{\xi}{}_{\mu}{}^{\kappa}
-\frac{1}{8}g_{\alpha\beta}g_{\mu\nu}R^{2}_{\rho\tau\lambda\sigma},\\
S_{\alpha\beta\mu\nu}&:=&R_{\alpha\mu\xi\kappa}R_{\beta\nu}{}^{\xi\kappa}
+R_{\alpha\nu\xi\kappa}R_{\beta\mu}{}^{\xi\kappa}
+\frac{1}{4}g_{\alpha\beta}g_{\mu\nu}R^{2}_{\rho\tau\lambda\sigma},\\
T_{\alpha\beta\mu\nu}&:=&-\frac{1}{8}(5g_{\alpha\beta}g_{\mu\nu}-g_{\alpha\mu}g_{\beta\nu}
-g_{\alpha\nu}g_{\beta\mu})R^{2}_{\rho\tau\lambda\sigma},\label{27bNov2018}
\end{eqnarray}
where $R_{\lambda\sigma\xi\kappa}$ is the Riemann tensor,
$R^{2}_{\rho\tau\lambda\sigma}=R_{\rho\tau\lambda\sigma}R^{\rho\tau\lambda\sigma}$,
Greek letters mean spacetime and the signature is $+2$. The
associated known energy-momentum density are
\begin{eqnarray}
B_{\mu{}0ij}\delta^{ij}=(E^{2}_{ab}+H^{2}_{ab},2\epsilon_{cab}E^{ad}H^{b}{}_{d}),~
S_{\mu{}0ij}\delta^{ij}=-\frac{2}{3}T_{\mu{}0ij}\delta^{ij}=-10(E^{2}_{ab}-H^{2}_{ab},\vec{0}),\label{29aSep2013}
\end{eqnarray}
where Latin denotes spatial indices, $E^{2}_{ab}=E_{ab}E^{ab}$ and
similarly for $H^{2}_{ab}$. The electric part $E_{ab}$ and
magnetic part $H_{ab}$, are defined in terms of the Weyl
curvature~\cite{Carmeli}: $E_{ab}:=C_{ambn}t^{m}t^{n}$ and
$H_{ab}:=*C_{ambn}t^{m}t^{n}$, where $t^{m}$ is the timelike unit
vector and $*C_{\rho\tau\xi\kappa}$ indicates its dual for the
evaluation. Here we emphasize that $B$ is completely trace free
which implies $B_{\mu000}=B_{\mu{}0ij}\delta^{ij}$. The energy
component of $B$ in (\ref{29aSep2013}) is non-negative definite
for all observers, i.e., positivity.  Meanwhile, the 4-momentum of
$B$ possesses the future directed non-spacelike property, i.e.,
\begin{eqnarray}
B_{0000}-|B_{000c}|\geq{}0.
\end{eqnarray}
In a small sphere limit, all of them satisfy the divergence free
condition:
$\partial_{\alpha}(\chi^{\alpha}{}_{\beta\mu\nu}x^{\mu}x^{\nu})$
is vanishing for all $\chi\in\{B,S,W\}$. This condition implies
the conservation of energy-momentum. In addition, we list the
following
\begin{eqnarray}
R^{2}_{\alpha\beta\mu\nu}=8(E^{2}_{ab}-H^{2}_{ab}),\quad{}
R_{\alpha\beta\mu\nu}*R^{\alpha\beta\mu\nu}=16E_{ab}H^{ab}.\label{30bDec2013}
\end{eqnarray}
There is a linear combination between $S$ and $T$
\begin{eqnarray}
W_{\alpha\beta\lambda\sigma}:=\frac{3}{2}S_{\alpha\beta\lambda\sigma}+T_{\alpha\beta\lambda\sigma},\label{27aNov2018}
\end{eqnarray}
such that it possesses zero energy-momentum, i.e.,
$W_{\mu{}000}=(0,0,0,0)$.

We observe that a certain kind of multiplication between $E_{ab}$
and $H_{ab}$ can be classified as the inner and cross products.
(i) Inner product: The momentum $E_{ab}H^{ab}$ can be expressed as
an analogy of an inner product
\begin{eqnarray}
E_{ab}H^{ab}&=&E_{1a}H^{1a}+E_{2a}H^{2a}+E_{3a}H^{3a}\nonumber\\
&=&\vec{E}_{1a}\cdot\vec{H}_{1b}+\vec{E}_{2a}\cdot\vec{H}_{2b}+\vec{E}_{3a}\cdot\vec{H}_{3b}\nonumber\\
&=&|\vec{E}_{1a}||\vec{H}_{1b}|\cos\theta_{1}\,\vec{n}_{1}
+|\vec{E}_{2a}||\vec{H}_{2b}|\cos\theta_{2}\,\vec{n}_{2}
+|\vec{E}_{3a}||\vec{H}_{3b}|\cos\theta_{3}\,\vec{n}_{3},
\end{eqnarray}
where $\theta_{1}$ is the angle between vectors
$(E_{11},E_{12},E_{13})$ and $(H_{11},H_{12},H_{13})$; similarly
for $\theta_{2}$ and $\theta_{3}$. Here we defined the
3-dimensional vector $\vec{E}_{1a}=(E_{11},E_{12},E_{13})$ and its
norm $|\vec{E}_{1a}|=\sqrt{E^{2}_{1a}}$. Similarly for
$\vec{E}_{2a}$ and $\vec{E}_{3a}$, etc.  It is possible for having
different representation since it is legitimated transform the
basis vectors from $\vec{i},\vec{j},\vec{k}$ to
$\vec{n}_{1},\vec{n}_{2},\vec{n}_{3}$. Here we consider the
absolute value
\begin{eqnarray}
|E_{ab}H^{ab}|&=&\sqrt{|\vec{E}_{1a}|^{2}|\vec{H}_{1b}|^{2}\cos^{2}\theta_{1}
+|\vec{E}_{2a}|^{2}|\vec{H}_{2b}|^{2}\cos^{2}\theta_{1}+|\vec{E}_{3a}|^{2}|\vec{H}_{3b}|^{2}\cos^{2}\theta_{3}}
\nonumber\\
&\leq&|\vec{E}_{1a}||\vec{H}_{1b}|+|\vec{E}_{2a}||\vec{H}_{2b}|+|\vec{E}_{3a}||\vec{H}_{3b}|.
\end{eqnarray}
(ii) Cross product: Consider another kind of momentum
\begin{eqnarray}
\epsilon_{cab}E^{ad}H^{b}{}_{d}
&=&(E_{2a}H^{3a}-E_{3a}H^{2a})\vec{i}+(E_{3a}H^{1a}-E_{1a}H^{3a})\vec{j}+(E_{1a}H^{2a}-E_{2a}H^{1a})\vec{k}\nonumber\\
&=&\left|
\begin{array}{ccccc}
\vec{i} & \vec{j} & \vec{k} \\
E_{11} & E_{12} & E_{13} \\
H_{11} & H_{12} & H_{13} \\
\end{array}
\right| +\left|
\begin{array}{ccccc}
\vec{i} & \vec{j} & \vec{k} \\
E_{21} & E_{22} & E_{23} \\
H_{21} & H_{22} & H_{23} \\
\end{array}
\right| +\left|
\begin{array}{ccccc}
\vec{i} & \vec{j} & \vec{k} \\
E_{31} & E_{32} & E_{33} \\
H_{31} & H_{32} & H_{33} \\
\end{array}
\right|\nonumber\\
&=&\vec{E}_{1a}\times\vec{H}_{1b}+\vec{E}_{2a}\times\vec{H}_{2b}+\vec{E}_{3a}\times\vec{H}_{3b}\nonumber\\
&=&|\vec{E}_{1a}||\vec{H}_{1b}|\sin\theta_{1}\vec{n}_{1}
+|\vec{E}_{2a}||\vec{H}_{2b}|\sin\theta_{1}\vec{n}_{2}
+|\vec{E}_{3a}||\vec{H}_{3b}|\sin\theta_{1}\vec{n}_{3},\label{20aNov2018}
\end{eqnarray}
The absolute magnitude for (\ref{20aNov2018}) can be manipulated
as
\begin{eqnarray}
|\epsilon_{cab}E^{ad}H^{b}{}_{d}|
=\sqrt{|\vec{E}_{1a}|^{2}|\vec{H}_{1b}|^{2}\sin^{2}\theta_{1}
+|\vec{E}_{2a}|^{2}|\vec{H}_{2b}|^{2}\sin^{2}\theta_{1}+|\vec{E}_{3a}|^{2}|\vec{H}_{3b}|^{2}\sin^{2}\theta_{3}}
\label{27aDec2018}\\
\leq|\vec{E}_{1a}||\vec{H}_{1b}|+|\vec{E}_{2a}||\vec{H}_{2b}|+|\vec{E}_{3a}||\vec{H}_{3b}|.
\quad\quad\quad\quad\quad\quad\quad\quad\quad\quad\quad\,
\label{21bDec2018}
\end{eqnarray}
Thus we have an identity
\begin{eqnarray}
|E_{ab}H^{ab}|^{2}+|\epsilon_{cab}E^{ad}H^{b}{}_{d}|^{2}
=E^{2}_{1a}H^{2}_{1b}+E^{2}_{2a}H^{2}_{2b}+E^{2}_{3a}H^{2}_{3b}.
\end{eqnarray}

The Bel-Robinson tensor possesses the dominate energy condition
which means it satisfies the future directed non-spacelike
property automatically. Perhaps, it may need to bear in mind that
$E_{ab}$ or $H_{ab}$ can be assigned any value. Here we illustrate
the future directed non-spacelike condition for the Bel-Robinson
`momentum':
\begin{eqnarray}
B_{0000}-|B_{000c}|&=&E^{2}_{ab}+H^{2}_{ab}-|2\epsilon_{cab}E^{ad}H^{b}{}_{d}|\nonumber\\
&\geq&(|\vec{E}_{1a}|-|\vec{H}_{1b}|)^{2}+(|\vec{E}_{2a}|-|\vec{H}_{2b}|)^{2}
+(|\vec{E}_{3a}|-|\vec{H}_{3b}|)^{2}\nonumber\\
&\geq&0,\label{31aOct2018}
\end{eqnarray}
where
$|B_{000c}|\leq{}2(|\vec{E}_{1a}||\vec{H}_{1b}|+|\vec{E}_{2a}||\vec{H}_{2b}|+|\vec{E}_{3a}||\vec{H}_{3b}|)$
which is indicated in (\ref{21bDec2018}). This result demonstrates
that how $B$ possesses the expected future directed non-spacelike
and indeed there does not exist any extra room to alter this
inequality at a first glance. More concretely, it is definitely
forbidden for adding any value of $(E^{2}_{ab}-H^{2}_{ab})$ or
$E_{ab}H^{ab}$. However, we claim there is a way out to make some
modification. Consider the following combination
\begin{eqnarray}
E^{2}_{ab}+H^{2}_{ab}-|2\epsilon_{cab}E^{ad}H^{b}{}_{d}|-|k_{1}||E^{2}_{ab}-H^{2}_{ab}|-|k_{2}||E_{ab}H^{ab}|
\geq{}0.\label{4aNov2018}
\end{eqnarray}
In order to make the above inequality holds, the unique solution
is when both constants $k_{1},k_{2}$ vanish simultaneous according
to Szabados suggested. This is called the pure Bel-Robinson
`momentum' requirement. Actually, we are repeating the same
argument with Szabados. To the contrary, we use another point of
view to examine the difference between the energy and momentum in
(\ref{31aOct2018}) again
\begin{eqnarray}
{\cal{E}}^{2}-|\vec{\cal{P}}|^{2}&=&(E^{2}_{ab}+H^{2}_{ab})^{2}-4(E^{2}_{1a}H^{2}_{1b}
+E^{2}_{2a}H^{2}_{2b}+E^{2}_{3a}H^{2}_{3b})\nonumber\\
&=&2(E^{2}_{1a}E^{2}_{2b}+E^{2}_{1a}E^{2}_{3b}+E^{2}_{2a}E^{2}_{3b}
+H^{2}_{1a}H^{2}_{2b}+H^{2}_{1a}H^{2}_{3b}+H^{2}_{2a}H^{2}_{3b})\nonumber\\
&&+2\left[E^{2}_{1a}(H^{2}_{2b}+H^{2}_{3b})
+E^{2}_{2a}(H^{2}_{1b}+H^{2}_{3b})
+E^{2}_{3a}(H^{2}_{1b}+H^{2}_{2b})
\right]\nonumber\\
&&+(E^{2}_{1a}-H^{2}_{1a})^{2}+(E^{2}_{2a}-H^{2}_{2a})^{2}+(E^{2}_{3a}-H^{2}_{3a})^{2}
\nonumber\\
&\geq&0,
\end{eqnarray}
where the momentum
$|\vec{\cal{P}}|\leq{}2\sqrt{E^{2}_{1a}H^{2}_{1b}+E^{2}_{2a}H^{2}_{2b}+E^{2}_{3a}H^{2}_{3b}}$
which is depicted in (\ref{27aDec2018}). Now adding the terms of
$(E^{2}_{ab}-H^{2}_{ab})$ or $E_{ab}H^{ab}$ are no longer
impossible. It turns out that we found a different result; one
that is strictly forbidden according to the conclusion of
Szabados's article. More precisely, what ranges for constants
$k_{1}$ and $k_{2}$ may be selected such that the future directed
non-spacelike qualities can be kept. For this purpose we use the
5-Petrov types Riemann curvatures for the
verification~\cite{Lobo}. We obtained the new constraints as
follows
\begin{eqnarray}
|k_{1}|\leq{}1,\quad{}2|k_{1}|+|k_{2}|\leq{}2.\label{3bDec2013}
\end{eqnarray}
This means the Bel-Robinson `momentum' is not an unique
energy-momentum that satisfies the future directed non-spacelike
requirement in a small sphere limit.

\section{Small sphere limit}
In a small sphere limit, we have proposed $V$ that contribute the
same pure Bel-Robinson `momentum' as $B$ does~\cite{SoPRD2009}.
The detail expression is $V=B+W$. Here we consider another
expression, a certain linear combination between $B$ with $S$, $T$
or $W$ in a small sphere limit. Note that, though the
energy-momentum for $W$ are vanishing, it is not zero for the
other components within this mentioned region. For example
$W_{0011}=2(E^{2}_{ab}-H^{2}_{ab}-3E^{2}_{1a}+3H^{2}_{1a})$.

Case (i): Consider a simple energy-momentum integral such that
within a small sphere limit, we consider a linear combination
between $B$ and $S$:
\begin{eqnarray}
B_{\alpha\beta\lambda\sigma}+a_{1}S_{\alpha\beta\lambda\sigma},
\end{eqnarray}
where $a_{1}$ is a real number. For constant time $t_{0}=0$, the
energy-momentum in vacuum with radius $r$
\begin{eqnarray}
2\kappa{\cal{}P}_{\mu}=\int_{t_{0}}(B^{0}{}_{\mu{}ij}+a_{1}S^{0}{}_{\mu{}ij})x^{i}x^{j}d^{3}x
=-\frac{4\pi}{15}r^{5}(B_{\mu{}000}+a_{1}S_{\mu{}0ij}\delta^{ij}),\label{5aOct2013}
\end{eqnarray}
where $\kappa=8\pi{}G/c^{4}$, $G$ is the Newtonian constant and
$c$ the speed of light. According to Szabados, the only
possibility is when $a_{1}=0$ that satisfies the positivity,
future pointing and non-spacelike properties~\cite{Szabados}.
Explicitly, the pure Bel-Robinson `momentum'. However, we claim
that there exists some non-vanishing $a_{1}$ such that these
future directed non-spacelike property is still preserved.
Referring to (\ref{29aSep2013}), we only vary the energy and
without affecting the momentum. Consequently, the energy-momentum
for (\ref{5aOct2013}) becomes
\begin{eqnarray}
({\cal{}E},{\cal{P}}_{c})=\frac{2\pi}{15\kappa}r^{5}
\left[(E^{2}_{ab}+H^{2}_{ab})-10a_{1}(E^{2}_{ab}-H^{2}_{ab}),2\epsilon_{cab}E^{a}{}_{d}H^{bd}\right].
\label{19aOct2013}
\end{eqnarray}
Generally, the values of $E_{ab}$ and $H_{ab}$ can be arbitrary at
a given point, the sign of the energy component of $S$ is
uncertain and obviously they should affect the future directed
non-spacelike property. Previously, our achievement preferred a
multiple of pure Bel-Robinson `momentum' in a small sphere region,
and we confidently sure that the result in (\ref{19aOct2013})
required $a_{1}$ vanishes~\cite{SoPRD2009}. Nevertheless, we found
a certain linear combinations of $B$ and $S$ are legitimate.
Referring to (\ref{20aNov2018}), we change another angle of view
for the comparison
\begin{eqnarray}
{\cal{}E}^{2}-|\vec{\cal{P}}|^{2}&\geq&\left[(E^{2}_{ab}+H^{2}_{ab})-10a_{1}(E^{2}_{ab}-H^{2}_{ab})\right]^{2}
-4(E^{2}_{1a}H^{2}_{1b}+E^{2}_{2a}H^{2}_{2b}+E^{2}_{3a}H^{2}_{3b})\nonumber\\
&\geq&0,\label{26aDec2018}
\end{eqnarray}
provided that $|a_{1}|\leq\frac{1}{40}$. Meanwhile, the
examination using the 5-Petrov types Riemann curvatures
verification serve a more precise value, i.e.,
$|a_{1}|\leq\frac{1}{10}$. Here we give a remark: previously we
believed both Einstein $t^{E}_{\alpha\beta}$ and Landau-Lifshitz
$t^{LL}_{\alpha\beta}$ pseudo-tensors could not pass the future
directed non-spacelike requirement in Riemann normal
coordinates~\cite{SoNesterCQG2009,Deser}:
\begin{equation}
t^{E}_{\alpha\beta}
=\frac{2}{9}\left(B_{\alpha\beta\xi\kappa}-\frac{1}{4}S_{\alpha\beta\xi\kappa}\right)x^{\xi}x^{\kappa},\quad{}
t^{LL}_{\alpha\beta}
=\frac{7}{18}\left(B_{\alpha\beta\xi\kappa}+\frac{1}{14}S_{\alpha\beta\xi\kappa}\right)x^{\xi}x^{\kappa}.
\label{15aMay2014}
\end{equation}
This illustration shows that we are mistaken in the past. Now, the
energy-momentum of the Landau-Lifshitz pseudo-tensor
(corresponding $|a_{1}|=\frac{1}{14}<\frac{1}{10}$) is a suitable
candidate for fulfilling the future directed non-spacelike
requirement. While Einstein pseudo-tensor still failed (associated
$|a_{1}|=\frac{1}{4}>\frac{1}{10}$).

Case (ii): Likewise, replace $S$ by $T$ which is indicated in
(\ref{27bNov2018}), the combination becomes $B+a_{2}T$. The future
directed non-spacelike property requires the constant
$|a_{2}|\leq\frac{1}{60}$ for a general comparison. The more
precise value, using the 5-Petrov types Riemann curvatures
examination, requires $|a_{2}|\leq\frac{1}{15}$. Summing up our
present result and the previous one, one can simply eliminate away
this extra energy and obtain the pure Bel-Robinson `momentum'.
This means there exists a linear combination, $S+\frac{2}{3}T$
which is denoted in (\ref{29aSep2013}) and (\ref{27aNov2018}),
contributes vanising energy-momentum.

Case (iii): Dominate energy condition in a small sphere limit.
Consider the energy-momentum stress in static
\begin{eqnarray}
t_{\alpha\beta}=\int{}t_{\alpha\beta{}ij}x^{i}x^{j}d^{3}x.
\end{eqnarray}
Suppose the energy $t_{00}=\int{}t_{00ij}x^{i}x^{j}d^{3}x$ is
positive definite. The requirement for the dominate energy
condition in a small sphere limit is $t_{00}\geq|t_{\alpha\beta}|$
for all $\alpha,\beta$. Here we consider the following combination
\begin{eqnarray}
B_{\alpha\beta\lambda\sigma}+s_{1}S_{\alpha\beta\lambda\sigma}+s_{2}W_{\alpha\beta\lambda\sigma},
\end{eqnarray}
where $s_{1},s_{2}$ are constants. The energy is
\begin{eqnarray}
\int(B_{00ij}+s_{1}S_{00ij}+s_{2}W_{00ij})x^{i}x^{j}d^{3}x\geq0,
\end{eqnarray}
provided that $|s_{1}|\leq\frac{1}{10}$ and the similar
manipulation can be found in (\ref{5aOct2013}). For a direct
comparison, let $s_{2}=10s_{1}$ and without using the 5-Petrov
types Riemann curvatures:
\begin{eqnarray}
\int\left[B_{00ij}+s_{1}(S_{00ij}+10W_{00ij})\right]x^{i}x^{j}d^{3}x
\geq\left|\int\left[B_{\alpha\beta{}ij}+s_{1}(S_{\alpha\beta{}ij}+10W_{\alpha\beta{}ij})\right]x^{i}x^{j}d^{3}x\right|,
\end{eqnarray}
is hold provided $|s_{1}|\leq\frac{1}{100}$ for all
$\alpha,\beta$. Hence, we have a simple combination that satisfies
the dominate energy condition in a small sphere limit. For the
completeness, as this combination expression possesses the
dominate energy condition, it must guarantee the future directed
non-spacelike property is hold. We verify this through the
following comparing
\begin{eqnarray}
\int\left[B_{00ij}+s_{1}(S_{00ij}+10W_{00ij})\right]x^{i}x^{j}d^{3}x
\geq\left|\int\left[B_{0cij}+s_{1}(S_{0cij}+10W_{0cij})\right]x^{i}x^{j}d^{3}x\right|,
\end{eqnarray}
and indeed it is hold when $|s_{1}|\leq\frac{1}{40}$. The same
explanation can be found after (\ref{26aDec2018}).

\section{Small ellipsoid}
Instead of integrate the energy-momentum in a small sphere limit,
one can consider small ellipsoid.  One of the natural options is
the Jupiter-Io system, Jupiter deforms Io from sphere to ellipsoid
through the tidal force and vice versa. Consider a simple
dimension $(a,b,c)=(\sqrt{1+\Delta},1,1)a_{0}$ for non-zero
$\Delta>-1$ and $a_{0}$ finite. In reality, it is slightly
deformed and it suits the quasi-local small 2-surface limit. The
physical dimension for Io is $(x,y,z)=(3660.0, 3637.4,3630.6)$ in
kilometer. Using our notation: $a=\sqrt{1+\Delta}\,a_{0}$,
$b\simeq{}c\simeq{}a_{0}$, where the mean radius $a_{0}=1817$ km
and $\Delta=0.0144$. This kind of deformation is called spheroid.
For constant time $t_{0}=0$, the corresponding 4-momentum are
\begin{equation}
2\kappa{\cal{}P}_{\mu}=\int_{t_{0}}t^{0}{}_{\mu{}ij}x^{i}x^{j}d^{3}x
=\frac{4\pi}{15}a^{5}_{0}\left[t^{0}{}_{\mu{}00}+\Delta{}t^{0}{}_{\mu{}11}\right]\sqrt{1+\Delta},
\label{21aJan2013}
\end{equation}
where $t^{0}{}_{\mu{}ij}$ can be replaced by $B$ or $S$. It may be
worthwhile to check what is the energy different in a small sphere
and ellipsoid limits. In other words, is there any energy change
from a small sphere deforms to ellipsoid? Here we use the
Schwarzchild metric in spherical coordinates~(see \S31.2 in
\cite{MTW}) for a simple test:
\begin{eqnarray}
ds^{2}=-\left(1-\frac{2M}{r}\right)dt^{2}+\left(1-\frac{2M}{r}\right)^{-1}dr^{2}
+r^{2}(dr^{2}+\sin^{2}\theta\,d\phi^{2}),
\end{eqnarray}
with the assumption that $Mr^{-1}<<1$, both the gravitational
constant $G$ and speed of light $c$ are unity. Certainly, there is
no momentum since we are dealing with a static spacetime. The
non-vanishing Riemann curvatures are
$R_{\hat{t}\hat{r}\hat{t}\hat{r}}=-R_{\hat{\theta}\hat{\phi}\hat{\theta}\hat{\phi}}
=-2Mr^{-3}$ and
$R_{\hat{t}\hat{\theta}\hat{t}\hat{\theta}}=R_{\hat{t}\hat{\phi}\hat{t}\hat{\phi}}
=-R_{\hat{r}\hat{\theta}\hat{r}\hat{\theta}}=-R_{\hat{r}\hat{\phi}\hat{r}\hat{\phi}}=Mr^{-3}$.
The energy for $B$ and $S$ are
\begin{eqnarray}
(B_{0000},B_{0011},S_{0000},S_{0011})=(6,-2,12,-28)\frac{M^{2}}{r^{6}},
\end{eqnarray}
where the value of the Kretschmann scalar
$R^{2}_{\lambda\sigma\rho\tau}=48M^{2}r^{-6}$.

Case (1): Referring to (\ref{21aJan2013}), replace $t$ by $B$, the
energy-momentum are
\begin{equation}
2\kappa{\cal{}P}_{\mu}=\frac{4\pi}{15}a^{5}_{0}\left[B^{0}{}_{\mu{}00}+\Delta{}B^{0}{}_{\mu{}11}\right]\sqrt{1+\Delta}.
\end{equation}
where
\begin{eqnarray}
B_{0011}=E^{2}_{ab}+H^{2}_{ab}-2E^{2}_{1a}-2H^{2}_{1a},\quad{}
B_{0c11}=2\epsilon_{cab}(E^{ad}H^{b}{}_{d}-2E^{a}{}_{1}H^{b}{}_{1})
\end{eqnarray}
This result alter the energy and momentum of $B_{\mu{}000}$
simultaneously, i.e., making it analogous with (\ref{3bDec2013}):
$k_{1}\neq{}0\neq{}k_{2}$. Verify the following quantities
\begin{eqnarray}
{\cal{E}}-|\vec{\cal{P}}|=B_{0000}+\Delta{}B_{0011}-|B_{000c}+\Delta{}B_{0c11}|.\label{21cDec2018}
\end{eqnarray}
Using the 5-Petrov types Riemann curvatures to compare the energy
and momentum in (\ref{21cDec2018}), the future directed
non-spacelike condition requires $\Delta\in(-1,3)$. Here we check
the energy different, from a small sphere deformed to ellipsoid,
for the Jupiter-Io system:
\begin{eqnarray}
{\cal{E}}_{\rm{sphere}}=6\left(\frac{4\pi{}M^{2}a^{5}_{0}}{15r^{6}}\right),\quad{}{\cal{E}}_{\rm{ellipsoid}}
=6.01\left(\frac{4\pi{}M^{2}a^{5}_{0}}{15r^{6}}\right)
\end{eqnarray}
where we have used $\Delta=0.0144$. These data indicated that the
small ellipsoid absorbs more energy than sphere.

Case (2): According to (\ref{21aJan2013}), replace $t$ by $B+S$,
the energy-momentum become
\begin{equation}
2\kappa{\cal{}P}_{\mu}
=\frac{4\pi}{15}\left[(B^{0}{}_{\mu{}ij}+sS^{0}{}_{\mu{}ij})\delta^{ij}
+\Delta(B^{0}{}_{\mu{}11}+sS^{0}{}_{\mu{}11})\right]
a^{5}_{0}\sqrt{1+\Delta},\label{14aMay2014}
\end{equation}
where $s$ is a constant. The energy-momentum for $S_{0\mu{}11}$
\begin{eqnarray}
S_{0011}&=&-2(E^{2}_{ab}-H^{2}_{ab}+2E^{2}_{1a}-2H^{2}_{1a}),\\
S_{0c11}&=&4(0,E_{1a}H_{3}{}^{a}+E_{3a}H_{1}{}^{a},-E_{1a}H_{2}{}^{a}-E_{2a}H_{1}{}^{a}).
\end{eqnarray}
Based on (\ref{15aMay2014}) for $s=\frac{1}{14}$ which means we
choose the Landau-Lifshitz pseudo-tensor as an illustration. Using
the 5-Petrov types Riemann curvatures for the verification, we
discovered that when $\Delta\in[-\frac{1}{3},\frac{1}{5}]$
satisfies the future directed non-spacelike requirement. Using the
similar technique in Case (1) to check the energy difference, from
small sphere deformed to ellipsoid. Here we focus on the
Jupiter-Io system as a simple demonstration
\begin{eqnarray}
{\cal{E}}_{\rm{sphere}}=1.71\left(\frac{4\pi{}M^{2}a^{5}_{0}}{15r^{6}}\right),\quad{}{\cal{E}}_{\rm{ellipsoid}}
=1.67\left(\frac{4\pi{}M^{2}a^{5}_{0}}{15r^{6}}\right)
\end{eqnarray}
where we have substituted $\Delta=0.0144$ again. These data
indicated that the small ellipsoid absorbs less energy than
sphere. Naively, apply to the Jupiter-Io tidal heating system
using the Landau-Lifshitz pseudo-tensor~\cite{Purdue}, one may
interpret that Io releases energy away when changing the shape
from sphere to ellipsoid. Meanwhile, Io absorbs more energy when
deforming the shape from ellipsoid to sphere. Eventually, Io does
not gain or lose any energy after a complete deformation cycle.
This indicates that the interior of Io is in thermal
equilibrium~\cite{LaineyNature}. Perhaps, this simple argument
might help a little understanding of the real physical situation
of Io.

\section{Conclusion}
To describe the positive quasi-local energy-momentum expression,
the Bel-Robinson tensor $B$ is the most ideal candidate because it
gives the Bel-Robinson `momentum' in a small sphere region. In the
past, it seems that only this Bel-Robinson `momentum' can manage
this specific task: non-spacelike and future pointing. That
particular restriction could not allow even a small amount of
energy-momentum to be subtracted from this Bel-Robinson
`momentum'. After some comparison, with or without the 5-Petrov
types Riemann curvatures, we discovered that the Bel-Robinson
`momentum' implies future directed non-spacelike properties; but
the converse is not true. In other words, the Bel-Robinson
`momentum' is no longer the unique option for achieving the future
pointing non-spacelike requirement. Explicitly, the Bel-Robinson
tensor lost its privilege. For example, we thought the
Landau-Lifshitz pseudo-tensor was failed meet the future directed
non-spacelike requirement, but now we find that it can.
Furthermore, we constructed a linear combination, $B$ with other
tensors $S$ and $T$, gives the dominate energy condition in a
small sphere limit.

Besides the Bel-Robinson tensor, there exists a certain relaxation
freedom such that one can still obtain the energy-momentum
expression contributes the future directed non-spacelike property.
For example, $B+sS$ in a small sphere and ellipsoid regions. This
comparison, in some sense, reflects the reality for the Jupiter-Io
system tidal heating through the shape changing of Io. More
precisely, the interior of Io is in the thermal equilibrium.

\end{document}